\documentclass[dvipdfmx,12pt]{article}
\usepackage[dvipsnames]{xcolor}
\usepackage{amsmath} 
\usepackage{amssymb}
\usepackage{mathtools}
\usepackage{physics}
\usepackage{bm}
\usepackage[margin =27truemm]{geometry}
 \usepackage{cite}
 \usepackage{color}
 
\begin{document}
\setlength{\baselineskip}{0.6cm}

\begin{titlepage}
\begin{flushright}
NITEP 202 
\end{flushright}

\vspace*{10mm}%

\begin{center}{\large\bf
Attempt at Constructing a Model of Grand Gauge-Higgs \\ 
\vspace*{2mm}
Unification with Family Unification 
}
\end{center}
\vspace*{10mm}
\begin{center}
{\large Nobuhito Maru}$^{a,b}$ and
{\large Ryujiro Nago}$^{a}$ 
\end{center}
\vspace*{0.2cm}
\begin{center}
${}^{a}${\it
Department of Physics, Osaka Metropolitan University, \\
Osaka 558-8585, Japan}
\\
${}^{b}${\it Nambu Yoichiro Institute of Theoretical and Experimental Physics (NITEP), \\
Osaka Metropolitan University,
Osaka 558-8585, Japan}
\end{center}

\vspace*{20mm}


\begin{abstract}
We discuss a possibility whether a model of grand gauge-Higgs unification 
incorporating family unification in higher dimensions can be constructed. 
We first extend a five dimensional $SU(6)$ grand gauge-Higgs unification model to 
a five dimensional $SU(7)$ grand gauge-Higgs unification model compactified on an orbifold $S^1/Z_2$ 
to obtain three generations of quarks and leptons after symmetry breaking of the larger family unified gauge group. 
A prescription of constructing a six dimensional $SU(N)$ grand gauge-Higgs unification model 
including a five dimensional $SU(7)$ grand gauge-Higgs unification 
after compactifying the sixth dimension on an orbifold $S^1/Z_2$ is given. 
We find a six dimensional $SU(14)$ grand gauge-Higgs unification model 
with a set of representations containing three generations of quarks and leptons. 
\end{abstract}

\end{titlepage}

\section{Introduction}
One of the mysteries in the Standard Model (SM) of particle physics is 
the origin of three generations of quarks and leptons, 
where quarks and leptons with the same representations and charges have a triple copy structure. 
Understanding whether three generations are inevitable or accidental is a very nontrivial problem. 
One of the approaches to solve this problem has been known as ``family unification". 
In this scenario, quarks and leptons are embedded into fermions (desirably one fermion) 
in some non-repetitive representations of a large group 
which has the grand unified theory (GUT) gauge group or the SM gauge group as a subgroup
and three generations of quarks and leptons appear after symmetry breaking. 
The study of family unification has a very long history and has been done from various viewpoints 
\cite{
Georgi, Frampton1, FN, Frampton2, IKK, WZ, NSS, Fujimoto, BDM, BSMR, Barr, FK1, BBK, 
HS, KKO, KK, FK2, KM, AFK, GKM, GK, RVVW, Yamatsu
}. 

A pioneering work was given by Georgi \cite{Georgi}. 
Three copies of fermions in $\overline{{\bf 5}}$ and ${\bf 10}$ of $SU(5)$ representations 
were obtained from some non-repetitive and anomaly-free set 
of totally anti-symmetric representations in $SU(11)$ theory. 
In a process of symmetry breaking, 
one might worry that there might appear many extra and unwanted massless fermions. 
If the representations of the fermions are vector-like or real under the unbroken subgroup, 
such fermions would have masses of the order of symmetry breaking scale and are decoupled in the low energy theory. 
Taking into account this point, three generations were obtained. 

This line of thought has also applied to the higher dimensional theory \cite{KKO}. 
The advantages of this application are as follows. 
In odd dimensional theories, the anomaly constraints for the representations of fermions are relaxed. 
This greatly helps the possibility of model building extended.   
In even dimensional theories, the anomaly constraints certainly exist in general, 
but it is not so serious if we consider only the Dirac fermions from the beginning. 
The other advantage is that it is relatively easy to consider symmetry breaking 
in higher dimensional theory compactified on an orbifold. 
In such theories, symmetry breaking can be realized by boundary conditions 
and we have no need to introduce a complicated and unnatural Higgs potential as in four dimensional theories. 

On the other hand, it has been known that the SM gauge fields and the SM Higgs field can be unified 
in higher dimensional gauge theory, which is called as ``gauge-Higgs unification". 
In the scenario, the SM Higgs field is identified with extra spatial components of the higher dimensional gauge field. 
Regardless of the non-renormalizability, 
the quantum corrections of the physical quantities as Higgs mass and potential are predicted to be finite, 
which solves the hierarchy problem. 
The GUT extension of this scenario have been also considered and are very interesting, 
which 
has been paid much attention 
\cite{LM, KTY, HosotaniGGHU, ABBGW}.  
One of authors has proposed a five dimensional (5D) $SU(6)$ grand gauge-Higgs unification (GGHU). 
It is remarkable in this model that quarks and leptons of one generation can be embedded into 
two $\overline{{\bf 6}}$ and single ${\bf 20}$ representations of $SU(6)$ without exotic fermions. 

In this paper, we attempt to incorporate the family unification into grand gauge-Higgs unification.   
Employing the above $SU(6)$ model, we immediately notice that we have to extend the $SU(6)$ model 
since the ${\bf 20}$ representation is self-conjugate to $\overline{{\bf 20}}$ 
and even if three ${\bf 20}$ are obtained, a pair of them makes a mass term, 
which implies that three generations are impossible in the approach of \cite{Georgi}. 
Therefore, we first extend the 5D $SU(6)$ GGHU model to an $SU(7)$ model to circumvent this problem. 
Then, we search for a six dimensional (6D) GGHU model with family unification 
following the approach \cite{Georgi}, which includes the 5D $SU(7)$ GGHU model with three generations 
after compactification to 5D. 
We find a 6D $SU(14)$ GGHU model with family unification. 

The organization of this paper is as follows. 
In section 2, 5D $SU(7)$ GGHU model is constructed as a simple extension of 5D $SU(6)$ GGHU model. 
In section 3, we investigate a GGHU model with family unification in 6D following the approach by \cite{Georgi}. 
We find an $SU(14)$ GGHU model with family unification. 
Conclusion is given in the last section. 

\section{5D $\bm{SU(7)}$ theory}
In this section, we point out some problems of our previous 5D $SU(6)$ GGHU model 
in the context of family unification and propose an $SU(7)$ model as a simple extension 
before discussing a 6D GGHU model of family unification. 

In \cite{LM}, a GGHU model of 5D $SU(6)$ gauge theory 
compactified on $S^1/Z_2$ was proposed,  
where the gauge fields and the Higgs fields in GUT are unified in the 5D gauge field. 
One of the remarkable features of this model is that one generation of quarks and leptons are embedded into 
a set of representations $2 \times \overline{\bm{6}} \oplus \bm{20}$ of $SU(6)$ 
and furthermore no massless exotic fermions are left after compactification. 
However, we immediately notice that three generations of quarks and leptons cannot be obtained 
from this $SU(6)$ model in the approach of \cite{Georgi}  
since the $\bm{20}$ is self-conjugate to $\overline{\bm{20}}$ under $SU(6)$ 
and therefore a pair of them should be massive. 

To improve this point, 
we propose a 5D $SU(7)$ model compactified on $S^1/Z_2$ including the $SU(6)$ model. 
$Z_2$ parity matrices are given on each fixed points $y=0, \pi R$ as
	\begin{equation}
		\begin{aligned}
			P_0 = \; & \mathrm{diag}( +1, +1, +1, +1, +1, -1, -1) \qq{at} y = 0, \\
			P_1 = \; & \mathrm{diag} (+1, +1, -1, -1, -1, -1, +1) \qq{at} y = \pi R,
		\end{aligned}
	\end{equation}
where $y, R$ are the coordinate of the fifth dimension and a radius of the $S^1$, respectively. 
For the gauge fields, the $Z_2$ parity boundary conditions are taken as follows. 
\begin{align}
A_\mu(x^\mu, -y) = P_{0,1} A_\mu (x^\mu, y)P_{0,1}^\dag, \quad
A_5(x^\mu, -y) = -P_{0,1} A_\mu (x^\mu, y)P_{0,1}^\dag. 
\end{align}
$x^\mu=0,1,2,3$ denotes the four dimensional spacetime coordinate. 

More explicitly, the parities of $A_{\mu, 5}$ can be expressed in a matrix form. 
\begin{align}
&A_\mu = \left(
\begin{tabular}{cc|ccc|c|c}
(+, +) & (+, +) & (+, $-$)& (+, $-$) & (+, $-$)& ($-, -$)& ($-, +$) \\
(+, +) & (+, +) & (+, $-$)& (+, $-$) & (+, $-$)& ($-, -$)& ($-, +$) \\
\hline
$(+, -)$ & $(+, -)$ & (+, $+$)& (+, $+$) & (+, $+$)& ($-, +$)& ($-, -$) \\
$(+, -)$ & $(+, -)$ & (+, $+$)& (+, $+$) & (+, $+$)& ($-, +$)& ($-, -$) \\
$(+, -)$ & $(+, -)$ & (+, $+$)& (+, $+$) & (+, $+$)& ($-, +$)& ($-, -$) \\
\hline
($-, -$) & ($-, -$) & ($-, +$)& ($-, +$) & ($-,+$)& (+, +) & ($+, -$) \\
\hline
($-, +$) & ($-, +$) & ($-, -$)& ($-, -$) & ($-, -$)& ($+, -$)& (+, +) \\
\end{tabular}
\right), \\
&A_5 = \left(
\begin{tabular}{cc|ccc|c|c}
$(-, -)$ & $(-, -)$ & ($-, +$)& ($-, +$) & ($-, +$)& ($+, +$)& ($+, -$) \\
$(-, -)$ & $(-, -)$ & ($-, +$)& ($-, +$) & ($-, +$)& ($+, +$)& ($+, -$) \\
\hline
$(-, +)$ & $(-, +)$ & (+, $-$)& (+, $-$) & (+, $-$)& ($+, -$)& ($+, +$) \\
$(-, +)$ & $(-, +)$ & (+, $-$)& (+, $-$) & (+, $-$)& ($+, -$)& ($+, +$) \\
$(-, +)$ & $(-, +)$ & (+, $-$)& (+, $-$) & (+, $-$)& ($+, -$)& ($+, +$) \\
\hline
($+, +$) & ($+, +$) & ($+, -$)& ($+, -$) & ($+, -$)& $(-, -)$ & ($-, +$) \\
\hline
($+, -$) & ($+, -$) & ($+, +$)& ($+, +$) & ($+, +$)& ($-, +$)& $(-, -)$ \\
\end{tabular}
\right)
\end{align}
where the matrix elements mean a set of $Z_2$ parities $( P_0, P_1 )$. 

Noting that Kaluza-Klein (KK) mode expansions of the five dimensional field $\Phi$ 
with $Z_2$ parity $( P_0, P_1 )$ are given by
	\begin{equation}
		\begin{aligned}
			\Phi^{(+,+)}(x,y) &= \frac{1}{\sqrt{2\pi R}}\phi_0(x) 
			+ \frac{1}{\sqrt{\pi R}}\sum_{n=1}^{\infty} \phi_n^{(+,+)}(x)\cos \qty(\frac{n}{R}\,y), \\
			\Phi^{(+,-)}(x,y) &= \frac{1}{\sqrt{\pi R}}\sum_{n = 0}^{\infty} \phi_n^{(+,-)}(x)\cos \qty(\frac{n + 1/2}{R}\,y), \\
			\Phi^{(-,+)}(x,y) &= \frac{1}{\sqrt{\pi R}}\sum_{n = 0}^{\infty} \phi_n^{(-,+)}(x)\sin \qty(\frac{n + 1/2}{R}\,y), \\
			\Phi^{(-,-)}(x,y) &= \frac{1}{\sqrt{\pi R}}\sum_{n = 0}^{\infty} \phi_n^{(-,-)}(x)\sin \qty(\frac{n}{R}\,y),
		\end{aligned}
	\end{equation}
$\phi_0(x)$ in $\Phi^{(+,+)}(x,y)$ only remain massless in 4D. 
From this observation, we find the symmetry breaking pattern 
$ SU(7) \to SU(3)_c \times SU(2)_L \times U(1)_Y \times U(1)_{\alpha} \times U(1)_{\beta} $ from $A_\mu$ 
and the SM Higgs doublet is indeed embedded in $A_5$. 
Generators of $U(1)_{\alpha}$ and $U(1)_{\beta}$ are contained in the $SU(7)$ generators as
	\begin{equation*}
		U(1)_{\alpha}:
			\begin{pmatrix}
				1 & {} & {} & {} & {} & {} & {} \\
				{} & 1 & {} & {} & {} & {} & {} \\
				{} & {} & 1 & {} & {} & {} & {} \\
				{} & {} & {} & 1 & {} & {} & {} \\
				{} & {} & {} & {} & 1 & {} & {} \\
				{} & {} & {} & {} & {} & -5 & {} \\
				{} & {} & {} & {} & {} & {} & 0 
			\end{pmatrix}, \quad
		U(1)_{\beta}:
			\begin{pmatrix}
				1 & {} & {} & {} & {} & {} & {} \\
				{} & 1 & {} & {} & {} & {} & {} \\
				{} & {} & 1 & {} & {} & {} & {} \\
				{} & {} & {} & 1 & {} & {} & {} \\
				{} & {} & {} & {} & 1 & {} & {} \\
				{} & {} & {} & {} & {} & 1 & {} \\
				{} & {} & {} & {} & {} & {} & -6
			\end{pmatrix}.
	\end{equation*}
Looking at $A_5$, we find the colored Higgs field to be massless at tree level. 
Taking into account quantum corrections, its mass squared is expected to be ${\cal O}(\alpha_s/R^2)$. 
Since a typical order of the compactification scale is ${\cal O}$(10TeV) in GHU, 
the colored Higgs mass will be at most ${\cal O}$(10TeV). 
For proton stability, the baryon number violating operators by the colored Higgs exchange 
have to be forbidden or suppressed enough by some mechanism, which will not be discussed in this paper.

In the $SU(7)$ theory, $2 \times \overline{\bm{6}} \oplus \bm{20}$ are embedded 
in $2 \times \overline{\bm{7}} \oplus \overline{\bm{35}}$:
	\begin{equation}
		\begin{aligned}
			\overline{\bm{7}} &= \overline{\bm{6}} \oplus \bm{1},\\
			\overline{\bm{35}} &= \bm{20} \oplus \overline{\bm{15}}.
		\end{aligned}
	\end{equation}
More explicitly, one generation of quarks and leptons are contained as follows
	\begin{equation}
		\begin{aligned}
		\overline{\bm{7}} &= \left\{ \begin{aligned}
				&\overline{\bm{7}}_L = \underbrace{ (\overline{\bm{3}}, \bm{1})^{(+,-)}_{ (\frac{1}{3}, -1, -1) }
								\oplus l_L (\bm{1}, \bm{2})^{(+,+)}_{ (-\frac{1}{2}, -1, -1) }
								\oplus(\bm{1}, \bm{1})^{(-,-)}_{ (0, 5, -1) } }_{\overline{\bm{6}}}
								\oplus \underbrace{(\bm{1}, \bm{1})^{(-,+)}_{ (0, 0, 6) } }_{\bm{1}} \\
				&\overline{\bm{7}}_R = \underbrace{ (\overline{\bm{3}}, \bm{1})^{(-,+)}_{ (\frac{1}{3}, -1, -1) }
								\oplus(\bm{1}, \bm{2})^{(-,-)}_{ (-\frac{1}{2}, -1, -1) }
								\oplus \nu_R (\bm{1}, \bm{1})^{(+,+)}_{ (0, 5, -1) } }_{\overline{\bm{6}}}
								\oplus \underbrace{ (\bm{1}, \bm{1})^{(+,-)}_{ (0, 0, 6) } }_{\bm{1}} \\
		\end{aligned} \right. \\
		\overline{\bm{7}} 
		&= \left\{ \begin{aligned}
				&\overline{\bm{7}}_L = \underbrace{ (\overline{\bm{3}}, \bm{1})^{(-,-)}_{ (\frac{1}{3}, -1, -1) }
								\oplus l_L (\bm{1}, \bm{2})^{(-,+)}_{ (-\frac{1}{2}, -1, -1) }
								\oplus(\bm{1}, \bm{1})^{(+,-)}_{ (0, 5, -1) } }_{\overline{\bm{6}}}
								\oplus \underbrace{ \chi_1 (\bm{1}, \bm{1})^{(+,+)}_{ (0, 0, 6) } }_{\bm{1}} \\
				&\overline{\bm{7}}_R = \underbrace{ \overline{d}_R (\overline{\bm{3}}, \bm{1})^{(+,+)}_{ (\frac{1}{3}, -1, -1) }
								\oplus(\bm{1}, \bm{2})^{(+,-)}_{ (-\frac{1}{2}, -1, -1) }
								\oplus(\bm{1}, \bm{1})^{(-,+)}_{ (0, 5, -1) } }_{\overline{\bm{6}}}
								\oplus \underbrace{ (\bm{1}, \bm{1})^{(-,-)}_{ (0, 0, 6) } }_{\bm{1}} \\
		\end{aligned} \right. \\
		\overline{\bm{35}} 
		&= \left\{ \begin{aligned}
			\overline{\bm{35}}_{L} = & \underbrace{ q_L(\bm{3}, \bm{2})^{(+,+)}_{ (\frac{1}{6}, -3, -3)}
								\oplus(\overline{\bm{3}}, \bm{1})^{(+,-)}_{(-\frac{2}{3}, -3, -3)}
								\oplus(\bm{1}, \bm{1})^{(+,-)}_{(1, -3, -3)} }_{\bm{20}} \notag \\
							& \hspace*{30mm} \underbrace{	
							\oplus(\overline{\bm{3}} ,\bm{2})^{(-,+)}_{(-\frac{1}{6}, 3, -3)}
								\oplus(\bm{3}, \bm{1})^{(-,-)}_{(\frac{2}{3}, 3., -3)}
								\oplus(\bm{1}, \bm{1})^{(-,-)}_{(-1, 3, -3)} }_{\bm{20}}\\
							& \oplus \underbrace{ (\overline{\bm{3}} ,\bm{2})^{(-,-)}_{(-\frac{1}{6}, -2, 4)}
								\oplus(\bm{3}, \bm{1})^{(-,+)}_{(\frac{2}{3}, -2., 4)}
								\oplus \chi_2 (\overline{\bm{3}}, \bm{1})^{(+,+)}_{(\frac{1}{3}, 4, 4)}
								\oplus(\bm{1} ,\bm{2})^{(+,-)}_{(-\frac{1}{2}, 4, 4)}
								\oplus(\bm{1}, \bm{1})^{(-,+)}_{(-1, -2, 4)}}_{\overline{\bm{15}}}\\
			\overline{\bm{35}}_{R} = & \underbrace{ (\bm{3}, \bm{2})^{(-,-)}_{ (\frac{1}{6}, -3, -3)}
								\oplus(\overline{\bm{3}}, \bm{1})^{(-,+)}_{(-\frac{2}{3}, -3, -3)}
								\oplus(\bm{1}, \bm{1})^{(-,+)}_{(1, -3)} }_{\bm{20}} \notag \\								
								& \hspace*{30mm} \underbrace{
								\oplus(\overline{\bm{3}} ,\bm{2})^{(+,-)}_{(-\frac{1}{6}, 3, -3)}
								\oplus u_R (\bm{3}, \bm{1})^{(+,+)}_{(\frac{2}{3}, 3, -3)}
								\oplus e_R (\bm{1}, \bm{1})^{(+,+)}_{(-1, 3, -3)}}_{\bm{20}}\\
							& \oplus \underbrace{ \chi_3 (\overline{\bm{3}} ,\bm{2})^{(+,+)}_{(-\frac{1}{6}, -2, 4)}
								\oplus(\bm{3}, \bm{1})^{(+,-)}_{(\frac{2}{3}, -2., 4)}
							      \oplus(\overline{\bm{3}}, \bm{1})^{(-,-)}_{(\frac{1}{3}, 4, 4)}
								\oplus(\bm{1} ,\bm{2})^{(-,+)}_{(-\frac{1}{2}, 4, 4)}
								\oplus(\bm{1}, \bm{1})^{(+,-)}_{(-1, -2, 4)}}_{\overline{\bm{15}}}\\
		\end{aligned} \right.
		\end{aligned}
	\end{equation}
where the bold face numbers in the right-hand side are the representations 
 under $SU(3)_c \times SU(2)_L$ and the numbers in the subscript are the charges 
 of $U(1)_Y \times U(1)_{\alpha} \times U(1)_{\beta}$.
 $L(R)$ means 4D left(right)-handed chiralities. 

Furthermore, we must note that these representations have exotic fermions 
$\chi_i (i = 1, 2, 3)$ absent in the Standard Model as the price of extending the gauge group $SU(6)$ to $SU(7)$.  
These exotic fermions can be massive and removed by introducing the 4D fermions $\overline{\chi}_i$ 
with conjugate representations and opposite chirality to $\chi_i$ 
and Dirac mass terms $m_i \overline{\chi}_i \chi_i$ on the fixed point. 
%
\section{6D $\bm{SU(N)}$ theory}
In this section, we attempt constructing a six dimensional (6D) $ SU(N) $ theory on $S^1/Z_2$ 
to realize a family unification of the previous 5D $ SU(7) $ theory.
Our strategy is based on an approach by Georgi \cite{Georgi} and 
its applications to higher dimensional theory by Kawamura et al \cite{KKO, KK, KM, GKM, GK}. 
In \cite{Georgi}, the 4D theory of the gauge group $SU(N)$ including $SU(5)$ as a subgroup 
with fermions in only antisymmetric tensor representation of $SU(N)$ was considered
and a possibility that three generations of quarks and leptons are obtained after symmetry breaking was explored. 
This is a very nontrivial requirement since all of antisymmetric tensor representations are not replicated 
and the fermion content has to be anomaly-free. 
Furthermore, if some fermions are vector-like or real representations under $SU(5)$ after symmetry breaking, 
they will have a mass of order of the symmetry breaking scale. 
In \cite{KKO, KK, KM, GKM, GK}, the argument by Georgi was applied to the higher dimensional theory. 
The advantages of this application are as follows.  
There is no anomaly-free conditions in the case of odd dimensions. 
If the Dirac fermions are considered,  the anomaly is cancelled even in even dimensions. 
Moreover, the symmetry breaking can be easily realized by boundary conditions in extra spatial dimensions. 
Following approaches by \cite{Georgi} and \cite{KKO, KK, KM, GKM, GK}, 
we would like to attempt constructing a 6D model 
where both gauge-Higgs unification and family unification are incorporated. 

In 6D theory, gamma matrices $ \Gamma^M \; (M = 0, 1, 2, 3, 5, 6) $ are $ 8 \times 8$ matrices 
satisfying Clifford algebra,
	\begin{equation}
		\acomm{\Gamma^M}{\Gamma^N} = 2 \eta^{MN} \quad (M, N = 0, 1, 2, 3, 5, 6),
	\end{equation}
where $\eta^{MN}$ is metric and $\eta = \mathrm{diag} ( +1, -1, -1, -1, -1, -1) $. 
In addition, $\Gamma^7$ is defined as
	\begin{equation}
		\Gamma^7 \coloneqq i \Gamma^0  \Gamma^1  \Gamma^2  \Gamma^3  \Gamma^5  \Gamma^6.
	\end{equation}
In our notation, the 6D Weyl fermions $\Psi_{\pm}$ are eigenstates of 
$\Gamma^7$ with eigenvalue $\pm1$, respectively and 
are decomposed into 4D Weyl fermions $\psi_L$ and $\psi_R$,
	\begin{equation}
		\Psi = \begin{pmatrix}
				\Psi_+ \\
				\Psi_-
			   \end{pmatrix}
			= \begin{pmatrix}
				\psi_{+L} \\
				\psi_{+R} \\
				\psi_{-L} \\
				\psi_{-R}
			   \end{pmatrix}. 
	\end{equation}
\par
By compactifying the sixth dimension $z$ to $ S^1/Z_2 $ orbifold with a radius $R'$ of $S^1$, 
we consider a symmetry breaking $ SU(N) \to SU(7) \times SU(p) \times SU(q) \times U(1)^2 $ 
realized by the following $ Z_2$ parities $ P'_0 $ and $ P'_1 $,
\begin{equation}
\begin{aligned}
P'_0 = \; & \mathrm{diag} \overbrace{(+1, \cdots, +1, \; +1, \cdots, +1, \; -1, \cdots, -1)}^{N} \qq{at} z = 0, \\
P'_1 = \; & \mathrm{diag}\underbrace{( +1, \cdots, +1}_{7}, \; \underbrace{-1, \cdots, -1}_{p}, \; 
\underbrace{+1, \cdots, +1)}_{q} \qq{at} z = \pi R',
\end{aligned}
\end{equation}
where the rank of the gauge symmetries are unchanged by the symmetry breaking, $ 7 + p + q = N $.  
We introduce here a symbol $[k]_N$, 
which means the rank $k$ totally antisymmetric tensor representation of $SU(N)$.
This is decomposed into multiplets of $ SU(7) \times SU(p) \times SU(q) $ as
\begin{equation}
		[ k ]_N = \sum_{ l = 0 }^{ k } \sum_{ m = 0 }^{ k - l } \sum_{ n = 0 }^{ k - l - m } \qty( [\,l\,]_7, [ m ]_p, [ n ]_q ).
\end{equation}
We note that $[k]_N $ does not exist for the case of $ k > N $. 
$[k]_N $ can be expressed as an antisymmetric part of the tensor product 
of $k$ fundamental representations $ \bm{N} $ :
\begin{equation}
		[k]_N = \qty(\bm{N} \times \cdots \times \bm{N} )_a
\end{equation}
where $a$ means an antisymmetric part. 
The $ Z_2 $ transformations at $ z = 0, \pi R $ of the 6D Dirac fermion 
in the $[k]_N$ representation are given as follows. 
\begin{equation}
	\begin{aligned}
	\qty(\bm{N} \times \cdots \times \bm{N} )_a & \to \eta_k \qty( P'_0 \bm{N} 
	     \times \cdots \times P'_0 \bm{N} )_a \qq{at} z = 0, \\
			\qty(\bm{N} \times \cdots \times \bm{N} )_a & \to  \eta'_k \qty( P'_1 \bm{N} 
			\times \cdots \times P'_1 \bm{N} )_a \qq{at} z = \pi R', 
      \end{aligned}
\end{equation}
where $ \eta_k, \eta'_k = \pm 1 $. 
For example, $ [1]_N = \bm{N} $ is decomposed as
	\begin{equation}
		\bm{N} = ( \bm{7}, \bm{1}, \bm{1} ) \oplus ( \bm{1}, \bm{p}, \bm{1} ) \oplus ( \bm{1}, \bm{1}, \bm{q} ),
	\end{equation}
and the $Z_2$ transformation at $z = 0$ can be read as
	\begin{equation}
		\bm{N} \to \eta_k P'_0 \bm{N} =  \eta_k ( \bm{7}, \bm{1}, \bm{1} ) 
		\oplus + \eta_k ( \bm{1}, \bm{p}, \bm{1} ) \oplus - \eta_k ( \bm{1}, \bm{1}, \bm{q} ).
	\end{equation}
In this way, we obtain the $ Z_2 $ parities $ \mathcal{P}'_0 $ 
and $ \mathcal{P}'_1 $ of the representation $ \qty( [\,l\,]_7, [ m ]_p, [ n ]_q ) $ 
	\begin{equation}
		\begin{aligned}
			\mathcal{P}'_0 &= ( -1)^n \eta_k = ( -1)^{ l + m } ( -1 )^k \eta_k, \\
			\mathcal{P}'_1 &= ( -1)^m \eta'_k = ( -1)^{ l + n } ( -1 )^k \eta'_k,
		\end{aligned}
	\end{equation}
where a condition $l+m+n=k$ is taken into account in the second quality.

Here we can define that ``LH'' fermion $ \Psi_- $ has $ \eta_k, \eta'_k $ factors, 
and 
``RH'' fermion $ \Psi_+ $ has  $ -\eta_k, -\eta'_k $ factors, 
more explicitly, 
\begin{align}
\Psi_- &:  ( -1)^{ l + m } ( -1 )^k \eta_k, \quad ( -1)^{ l + m } ( -1 )^k \eta'_k, \\
\Psi_+ &:  -( -1)^{ l + m } ( -1 )^k \eta_k, \quad -( -1)^{ l + m } ( -1 )^k \eta'_k. 
\end{align}
This relative sign is due to that of the chiral operator in 6D. 
\par
Now choosing as $ \qty( (-1)^k \eta_k,  (-1)^k \eta'_k ) = (+1, +1) $ for simplicity,  
we obtain the $Z_2$ parity assignment for ${\bf 7}, {\bf 21}$ and ${\bf 35}$ representations 
of $SU(7)$ by putting $l=1, 2, 3$ as in the following Table 1. 
	\begin{table}[h]
	\begin{center}
		\begin{tabular}{c|c|c} \hline
			{} & $\mathcal{P}'_0$ & $\mathcal{P}'_1$ \\ \hline \hline
			$\bm{7}_{\pm}$ & $\pm(-1)^m$ & $\pm(-1)^n$ \\
			$\bm{21}_{\pm}$ & $\mp(-1)^m$ & $\mp(-1)^n$ \\
			$\bm{35}_{\pm}$ & $\pm(-1)^m$ & $\pm(-1)^n$ \\
			$\overline{\bm{35}}_{\pm}$ & $\mp(-1)^m$ & $\mp(-1)^n$ \\
			$\overline{\bm{21}}_{\pm}$ & $\pm(-1)^m$ & $\pm(-1)^n$ \\
			$\overline{\bm{7}}_{\pm}$ & $\mp(-1)^m$ & $\mp(-1)^n$ \\
		\end{tabular}
		\caption{$Z_2$ parity assignment of ${\bf 7}, {\bf 21}$ 
		and ${\bf 35}$ (and their conjugate) representations of $SU(7)$.
		$\pm$ are given by the corresponding $\Gamma_7$ eigenvalues.}
			\end{center}
	\end{table} \\
%
From the information of $Z_2$ parities listed in Table 1, 
we can obtain the numbers of massless fermions with $Z_2$ even parity in the representations of 
$ \bar{\bm{7}}_{-}, \bm{21}_{-} $ and $\overline{\bm{35}}_{-}$ 
\begin{equation}
		\begin{aligned}
			n^{(+,+)}_{\overline{\bm{7}}_{-}, k } &= \sum_{ m, n = \mathrm{even} } {}_p C_m \cdot {}_q C_n 
			\cdot \qty( \delta_{ k, 1 + m + n } + \delta_{ k, 6 + m + n } ) \\
	& \hspace{2cm} - \sum_{ m, n = \mathrm{odd} } {}_p C_m \cdot {}_q C_n \cdot \qty( \delta_{ k, 1 + m + n } 
	+ \delta_{ k, 6 + m + n } ), \\
			n^{(+,+)}_{\bm{21}_{-}, k } &= \sum_{ m, n = \mathrm{even} } {}_p C_m \cdot {}_q C_n 
			\cdot \qty( \delta_{ k, 2 + m + n } + \delta_{ k, 5 + m + n } ) \\
				& \hspace{2cm} - \sum_{ m, n = \mathrm{odd} } {}_p C_m \cdot {}_q C_n 
				\cdot \qty( \delta_{ k, 2 + m + n } + \delta_{ k, 5 + m + n } ), \\
			n^{(+,+)}_{\overline{\bm{35}}_{-}, k } &= \sum_{ m, n = \mathrm{even}} {}_p C_m 
			\cdot {}_q C_n \cdot \qty( \delta_{ k, 3 + m + n } + \delta_{ k, 4 + m + n } ) \\
				& \hspace{2cm} - \sum_{ m, n = \mathrm{odd}} {}_p C_m \cdot {}_q C_n 
				\cdot \qty( \delta_{ k, 3 + m + n } + \delta_{ k, 4 + m + n } )
		\end{aligned}
		\label{zeromodes}
\end{equation}
where $ \delta_{ k, l + m + n }$ expresses $ k = l + m + n $. 
For the case of $ \bar{\bm{7}}_{-}$, 
the net number of massless 5D fermions is given by 
the difference between the sum with respect to even $m, n$ come 
from $\bar{\bm{7}}_-$, $\bm{7}_+$ and that with respect to odd $m, n$ from $\bm{7}_-$, $\bar{\bm{7}}_+$), 
\begin{align}
n_{\bar{\bm{7}}_{-}} = \#\bar{\bm{7}}_{-} + \#\bm{7}_{+} - \#\bar{\bm{7}}_{+} - \#\bm{7}_{-}
\end{align}
where \# expresses the number of each multiplet.
Similar arguments for  $\bm{21}_{-}$ and $ \overline{\bm{35}}_{-}$ are also applied. 

Now, we would like to find a 6D family unified model of 5D SU(7) grand gauge-Higgs unification 
with three replicated $2 \times \overline{\bm{7}} \oplus \overline{\bm{35}}$ representations 
including three families of quarks and leptons. 

Taking $ N = 14, \, p = 5, $ and $q = 2$ of our interest, (\ref{zeromodes}) is rewritten as
\begin{equation}
\begin{aligned}
    n^{(+,+)}_{\overline{\bm{7}}_{-}, k } &= \sum_{ m = 0, 2, 4 } \sum_{ n = 0 , 2 } 
    {}_5 C_m \cdot {}_2 C_n \cdot \qty( \delta_{ k, 1 + m + n } + \delta_{ k, 6 + m + n }  ) \\
	& \hspace{2cm} - \sum_{ m = 1, 3, 5 } \sum_{ n = 1} {}_5 C_m 
	\cdot {}_2 C_n \cdot \qty( \delta_{ k, 1 + m + n } + \delta_{ k, 6 + m + n } ), \\
			n^{(+,+)}_{\mathbf{21}_{-}, k } &= \sum_{ m = 0, 2, 4 } \sum_{ n = 0 , 2 } {}_5 C_m 
			\cdot {}_2 C_n \cdot \qty( \delta_{ k, 2 + m + n } + \delta_{ k, 5 + m + n } ) \\
       & \hspace{2cm} - \sum_{ m = 1, 3, 5 } \sum_{ n = 1} {}_5 C_m 
       \cdot {}_2 C_n \cdot \qty(  \delta_{ k, 2 + m + n } + \delta_{ k, 5 + m + n } ),\\
   n^{(+,+)}_{\overline{\bm{35}}_{-}, k } &= \sum_{ m = 0, 2, 4 } \sum_{ n = 0 , 2 } {}_5 C_m 
   \cdot {}_2 C_n \cdot \qty(\delta_{ k, 3 + m + n } + \delta_{ k, 4 + m + n } ) \\
	& \hspace{2cm} - \sum_{ m = 1, 3, 5 } \sum_{ n = 1} {}_5 C_m \cdot {}_2 C_n 
	\cdot \qty( \delta_{ k, 3 + m + n } + \delta_{ k, 4 + m + n }). 
\end{aligned}
\end{equation}
From these results, the list of $ n^{(+,+)}_{\overline{\bm{7}}_-, k}, n^{(+,+)}_{\bm{21}_-, k} $ 
and $ n^{(+,+)}_{\overline{\bm{35}}_-, k}$ for various values of $k$ in Table 2 are obtained.
	\begin{table}[h]
	\begin{center}
		\begin{tabular}{c|c|c|c} \hline 
			$k$  & $n^{(+,+)}_{\overline{\bm{7}}_{-}, k }$ & $n^{(+,+)}_{\bm{21}_{-}, k}$ 
			& $n^{(+,+)}_{\overline{\bm{35}}_{-}, k }$ \\ \hline \hline
			1  & 1 & 0 & 0 \\
			2  & 0 & 1 & 0 \\
			3  & 1 & 0 & 1 \\
			4  & 0 & 1 & 1 \\
			5  & $-5$ & 1 & 1 \\
			6  & 1 & $-5$ & 1 \\
			7  & 3 & 1 & $-5$ \\
			8  & 1 & 3 & $-5$ \\
			9  & 0 & $-5$ & 3 \\
			10 & $-5$ & 0 & 3 \\
			11 & 0 & 3 & 0 \\
			12 & 3 & 0 & 0 \\
			13 & 0 & 0 & 0 
		\end{tabular}
		\caption{Numbers of massless fermions in 
		$\overline{\bm{7}}_{-}, \bm{21}_{-}, \overline{\bm{35}}_{-}$ representations 
		for various $k$-rank totally anti-symmetric tensors of $SU(14)$. }	
		\end{center}
	\end{table}
We note that $ n^{(+,+)}_{R_{-}, k } = -5 $ means $n^{(+,+)}_{R_{+}, k } = 5 $ 
because of $ n^{(+,+)}_{R_-, k } = - n^{(+,+)}_{R_{+}, k } ( R = \overline{\bm{7}}, \bm{21}, \overline{\bm{35}} )$.
Consider a set of representations $ [1]_{14} \oplus [2]_{14} \oplus [3]_{14}  
\oplus [4]_{14}  \oplus [6]_{14}  \oplus [11]_{14}  \oplus [12]_{14} $, 
all numbers of $\bar{\bm{7}}_{-}, \bm{21}_{-} $ and $\overline{\bm{35}}_{-}$ are given by
	\begin{equation}
		\begin{aligned}
		n_{\overline{\bm{7}}_{-}} & \coloneqq \sum_{k = 1, 2, 3, 4, 6, 11, 12} n^{(+,+)}_{\overline{\bm{7}}_{-}, k} 
									= 1 + 0 + 1 + 0 + 1+ 0 + 3 = 6, \\
		n_{\bm{21}_{-}} & \coloneqq \sum_{k = 1, 2, 3, 4, 6, 11, 12} n^{(+,+)}_{\mathbf{21}_{+}, k} 
							 = 0 + 1 + 0 + 1 - 5 + 3 + 0 = 0, \\
		n_{\overline{\bm{35}}_{-}} & \coloneqq \sum_{k = 1, 2, 3, 4, 6, 11, 12} n^{(+,+)}_{\overline{\bm{35}}_- ,k} 
									  = 0 + 0 +  1 + 1 + 1+ 0 + 0 = 3. 
		\end{aligned}
	\end{equation}
Remarkably, we have succeeded in obtaining a desirable set of massless fermions 
in the representations $3 \times (2 \times \overline{\bm{7}} \oplus \overline{\bm{35}}) $, 
which contain three families of quarks and leptons. 
We think this result to be very nontrivial 
since the number of massless $\bm{21}$ representations has to vanish 
in addition to $3 \times (2 \times \overline{\bm{7}} \oplus \overline{\bm{35}}) $ to realize three generations. 
We have investigated various models up to $SU(14)$, the above $SU(14)$ model was an only satisfactory solution. 
One might think that other models can obtained if larger gauge groups are considered. 
However, it seems to be difficult to find other GGHU models with family unification 
since the representations and the number of massless fermions become more complicated in such cases.  
It is therefore plausible to conclude that our 6D $SU(14)$ theory obtained in this paper 
is a simplest GGHU model with family unification in a class of $SU$ theories 
although we cannot prove it formally.  

In our model of GGHU with family unification, 
it is a very nontrivial issue to obtain Yukawa coupling. 
We note that the left-handed quark doublet and the right-handed up quark are 
included in $\overline{35}$ representation of the 5D fermion, 
which come from $[3]_{14}, [4]_{14}, [6]_{14}$ representations of 6D SU(14) theory. 
This means that up-type Yukawa coupling can be obtained 
from the gauge interaction in extra spatial components of SU(14). 
However, the right-handed down quark are included in $\overline{7}$ represention 
 different from $\overline{35}$ representation where the left-handed quark doubled is present. 
Similarly, the representations where the left-handed lepton doublet and the right-handed electron 
are included are different. 
Therefore, the down-type quark and the charged lepton Yukawa couplings cannot be obtained 
from the SU(14) gauge interactions such as the up-type quark Yukawa couplings. 
One of the setups to realize the down-type and the charged Yukawa couplings has been known 
that 4D quarks and leptons are introduced on the brane and they couple to the bulk fermions 
through the localized Dirac mass terms. 
Then, integrating out the bulk fermions provides non-local Yukawa coupling, 
which was proposed in \cite{SSS} and has been extensively studied in SU(6) GGHU model \cite{MYT}. 
We also have to construct such a setup in the present model along the above idea, 
but it seems to be very nontrivial and complicated, which is beyond the scope of this paper.  
 
Some comments on anomaly cancellation are given. 
A chiral gauge theory in even dimensions, 
in general, is possible to have gauge anomalies and global anomaly.  
Fermions in the representations $ [k]_{14} \; ( k = 1, 2, 3, 4, 6, 11, 12)$ we introduced 
are massless 6D Dirac fermions, that is, both of $ [k]_{14+} $ and $ [k]_{14-} $ are introduced.
Therefore, the gauge anomalies are canceled.
As for the global anomaly, 
the global anomaly is absent 
since the sixth homotopy group of $SU(14)$, $\Pi_6(SU(14))$, is known to be vanished \cite{DP}. 

\section{Conclusion}

We have attempted constructing a grand gauge-Higgs unification model with family unification in this paper.
A 6D $SU(14)$ GGHU theory as a family unified 5D $SU(7)$ GGHU theory was found, 
where the set of representations 
$ [1]_{14} \oplus [2]_{14} \oplus [3]_{14}  \oplus [4]_{14}  \oplus [6]_{14}  \oplus [11]_{14}  \oplus [12]_{14} $ 
are decomposed into $3 \times (2 \times \overline{\bm{7}} \oplus \overline{\bm{35}}) $ 
after symmetry breaking $SU(14) \to SU(7) \times SU(5) \times SU(2) \times U(1)^2 $.
Three generations of quarks and leptons are obtained from 
$3 \times (2 \times \overline{\bm{7}} \oplus \overline{\bm{35}}) $ after compactification to 4D. 
As far as we know, this is the first GGHU model unifying three generations of quarks and leptons, 
which is expected to be a guideline of a model building along this line.  

\par
We have many issues to be explored since our work done in this paper is still a first step 
towards a construction of realistic model of GGHU with family unification. 

In our model, we have unfortunately some massless exotic fermions. 
It is of course desirable to find a model without them.  
In order to realize it, it would be interesting to study the theories 
with other gauge groups and fermion matter content.   
In this paper, we have set $ \qty( (-1)^k \eta_k,  (-1)^k \eta'_k ) = (+1, +1) $ for all $k$ for simplicity.
Considering other patterns for $\qty( (-1)^k \eta_k,  (-1)^k \eta'_k )$ might help 
reducing the number of massless exotic fermions. 

In our family unified 6D $SU(14)$ model, two step compactification is considered. 
It would be interesting to investigate a possibility that three generations of quarks and leptons 
are directly obtained from the compactification of 6D GGHU theory to 4D one such as \cite{GKM}. 

As one of the other interesting directions, it would be interesting to consider models of $SO$ gauge theory. 
The family unification via the spinor representation in $SO$ theory has been much studied since the work \cite{WZ}. 
Although several irreducible representations are necessary to realize the family unification in $SU$ theory, 
the family unification in $SO$ theory is expected to be realized 
by fewer irreducible representations than $SU$ theory 
since the dimension of spinor representation is exponentially large with respect to the size of the gauge group.

There are many phenomenological issues to be studied, 
generating the realistic Yukawa hierarchy, the study of electroweak symmetry breaking, the flavor physics, 
the gauge coupling unification and the proton decay analysis and so on. 

All of them are left for our future study.




\end{document}